\documentclass[doublecol]{epl2}
\usepackage{amsmath}
\usepackage{amsfonts}
\usepackage{graphicx}
\usepackage{dcolumn}
\usepackage{bm}

\title{Quantum chaotic system as a model of decohering environment}
\author{Jayendra N. Bandyopadhyay\footnote{Present address : Department of Physics, National
University of Singapore, 117542, Singapore}}
\institute{Max-Planck Institute for the Physics of Complex Systems, 
N\"{o}thnitzerstr. 38, D-01187 Dresden, Germany}


\abstract{As a model of decohering environment, we show that quantum chaotic system behave 
equivalently as many-body system. An approximate formula for time-evolution of the reduced 
density matrix of a system interacting with a quantum chaotic environment is derived. This 
theoretical formulation is substantiated by numerical study of decoherence of two qubits 
interacting with a quantum chaotic environment modeled by chaotic kicked top. Like many-body 
model of environment, quantum chaotic system is efficient decoherer, and it can generate 
entanglement between the two qubits which have no direct interaction.}

\begin{document}

\maketitle

Interaction of a quantum system with environment creates correlations between the states 
of the system and of the environment. These correlations destroy the superposition of the 
system states - a phenomenon known as decoherence \cite{zurek_RMP}. This phenomenon is 
believed to be responsible for quantum to classical transition. Decoherence is also a major 
obstacle for designing quantum computational and informational protocol \cite{quant_comp}. 
Therefore, a deeper understanding of the phenomenon is required to address the fundamental 
question like quantum-classical transition and, to develop quantum computational protocol.

In general, the environment is modeled by many-body system, e.g. infinitely many harmonic 
oscillators in thermal equilibrium (Feynman-Vernon or Caldeira-Leggett model) 
\cite{feynman}, spin-boson model \cite{leggett}, chaotic spin-chain \cite{lages}, etc. In 
another approach, random matrix model of the environment is used 
\cite{rmt_model1,rmt_model2}. Random matrix theory has well known connections with quantum 
chaotic systems. Hence, some studies have concentrated on the possibility of having quantum 
dissipation and decoherence due to the interaction with chaotic degrees of freedom 
\cite{doron}. Recently, as a model of decohering environment, single particle quantum 
chaotic system has been considered \cite{casati}. This paper shows that the kicked rotator, 
a well studied model of chaotic system, can reproduce the decohering effects of a many-body 
environment. In comparison to complex many-body system, this simple deterministic system is 
very convenient for numerical as well as analytical studies of decoherence. Hence, single 
particle quantum chaotic system warrants a special attention as a model of decohering 
environment.

In this Letter, we establish direct equivalence of single particle quantum chaotic 
environment and Caldeira-Leggett type model of many-body environment by providing a rigorous 
but straightforward treatment. We keep our results vis-a-vis a recent study in which 
decoherence in a quantum system is investigated under the influence of a collection of 
harmonic oscillators environment \cite{braun_haake}. Our derivation first assumes weak 
interaction between the system and the environment. Using the interaction strength as a 
small parameter, we perform perturbative theory calculation. By exponentiating the 
perturbative expansion, we get an approximate formula for the non-perturbative strong 
interaction effect of the environment on the system. The approximate formula is then 
justified by numerical evidences. In numerics, we study decoherence of two noninteracting 
qubits which are individually interacting with a common quantum chaotic environment. We use 
chaotic kicked top, a very well studied model of quantum chaotic system \cite{haake_book}, 
as the environment.

Most general form of the Hamiltonian of a system $S$, interacting with an environment $E$, 
is $H = H_S + H_E + H_{SE}$, where $H_S$ and $H_E$ are the Hamiltonians of the system and 
the environment, respectively, and $H_{SE}$ is the system-environment coupling Hamiltonian.  
We assume throughout this Letter that the decoherence time is much smaller than the system 
characteristic time. Hence we can neglect any dynamics of the isolated system and can 
discard $H_S$. We consider kicked quantum chaotic system as a model of the environment, so 
the general form of the time-dependent system-environment Hamiltonian is : $H(t) = I_S 
\otimes H_E(t) + H_I(t)$ where $H_E(t) = H_1 + H_2 \sum_n \delta(t-n)$ and $H_I(t) = \alpha 
V_S \otimes V_E$; $V_S$ and $V_E$ are the coupling agents of the system 
and the environment, respectively. Parameter $\alpha$ determines the strength of the 
interaction. We now assume that $V_E$ commutes with $H_1$. This is a very natural assumption
as it separates environment dynamics from the system-environment interaction. The corresponding 
time-evolution operator $U$ in between two consecutive kicks 
is: $U=U_I U_0 = U_I (I_S \otimes U_E)$, where the coupling part $U_I = \exp(-i \alpha V_S 
\otimes V_E)$, $I_S$ is a unit matrix which indicates the absence of any dynamics in the 
system, and $U_E = \exp(-i H_1) \exp(-i H_2)$. Further, we assume that the initial joint 
state of the system-environment is an unentangled pure state of the form 
$|\Psi_{SE}(0)\rangle = |\psi_S(0)\rangle \otimes |\psi_E(0)\rangle$. We measure decoherence 
of the system by its loss of purity $P(n) \equiv \mbox{Tr}_S[\rho_S(n)^2]$ which varies from 
$1$, for pure state, to $1/N_S$ for completely mixed state, $N_S$ being the Hilbert space 
dimension of $S$. And $\rho_S(n) \equiv \mbox{Tr}_E |\Psi_{SE}(n)\rangle \langle 
\Psi_{SE}(n)|$ is the system reduced density matrix (RDM), where $|\Psi_{SE}(n)\rangle$ is 
the joint state of the system and the environment at time $n$.

We are interested to study entanglement between the two non-interacting qubits due to their 
interaction with a common chaotic environment. Following Wootters, the entanglement is measured 
by computing the concurrence
$C[\rho_S(n)] = \max(\Lambda,\,0)$, where $\Lambda \equiv \sqrt{\lambda_1}-\sqrt{\lambda_2}-
\sqrt{\lambda_3}-\sqrt{\lambda_4}$ and $\lambda_i$'s are the eigenvalues of the matrix 
$R(n)=\rho_S(n) \widetilde{\rho_S(n)}$; $\widetilde{\rho_S(n)} \equiv (\sigma_y \otimes 
\sigma_y) \rho_S(n)^\star (\sigma_y \otimes \sigma_y)$, $\rho_S(n)^\star$ being the complex 
conjugation of $\rho_S(n)$ in the computational basis $\{|00\rangle, |01\rangle, 
|10\rangle, |11\rangle\}$ \cite{wootters}. The concurrence varies from $0$, for separable state, 
to $1$ for maximally entangled state.

Our perturbation theory approach is reminiscent of the method followed by Tanaka et. al. in 
the context of entanglement production between two coupled chaotic systems \cite{tanaka}. We 
define the interaction picture of the joint density matrix $\overline{\rho_{SE}}(n) \equiv 
U_0(n)^\dagger \rho_{SE}(n) U_0(n)$, and of any arbitrary operator $O(n) \equiv 
U_0(n)^\dagger O U_0(n)$, where $U_0(n) \equiv (I_S \otimes U_E)^n$, and $O(n)$ represents 
the free evolution of $O$. Thus, the time evolution of $\overline{\rho_{SE}}(n)$ is 
determined by the mapping : $\overline{\rho_{SE}}(n) = U_I(n) \overline{\rho_{SE}}(n-1) 
U_I(n)^\dagger$. The perturbative expansion for $U_I(n)$ is :
\begin{equation}
U_{I}(n) = 1 - i \alpha V(n) - \frac{1}{2} \alpha^2 V(n)^2 + \mathcal{O} (\alpha^3),
\end{equation}
where $V(n) \equiv V_S(n) \otimes V_E(n)$, $V_S(n)$ and $V_E(n)$ are as usual free 
evolutions of $V_S$ and $V_E$, respectively. We introduce the eigenvalues and the 
eigenvectors of $V_S$ as $V_S |s\rangle = \lambda_s |s\rangle$. By tracing out the 
environment, we obtain perturbative expansion of RDM of the system $S$ at time $n$ 
in the eigenbasis of $V_S$ upto $\mathcal{O} (\alpha^2)$ as :
\begin{align}
\label{main1}
\left[\rho_S(n)\right]_{ss^\prime} \simeq \left[1 - i\, \alpha (\lambda_s-\lambda_{s^\prime}) 
g(n) -\frac{1}{2} \alpha^2 (\lambda_s-\lambda_{s^\prime})^2 g(n)^2 \right. \nonumber\\ 
-\alpha^2 (\lambda_s-\lambda_{s^\prime})^2 f(n) \biggl. - 
i\, \alpha^2 (\lambda_s^2 - \lambda_{s^\prime}^2) \phi(n) \biggl] 
\left[\rho_S(0)\right]_{ss^\prime},  
\end{align}
where $g(n) \equiv  \sum_{l=1}^n \langle V_E(l) \rangle$, and
\begin{eqnarray}
\label{def_fn}
f(n) &\equiv & \mbox{Re}\, \Phi(n)~;~\phi(n) \equiv \mbox{Im}\, \Phi(n),\nonumber\\
\Phi(n) &\equiv & \frac{1}{2} \sum_{l=1}^n \Bigl\langle V_E(l)^2 \Bigr\rangle + \sum_{l=2}^n 
\sum_{l^\prime=1}^{l-1} \Bigl\langle V_E(l) V_E(l^\prime) \Bigr\rangle\nonumber\\ 
&-& \frac{1}{2} \left(\sum_{l=1}^n \Bigl\langle V_E(l) \Bigr\rangle \right)^2.   
\end{eqnarray}
Here, $\langle \, \bullet\, \rangle \equiv \mbox{Tr}\{\rho_E(0)\, \bullet\}$ is an average 
over the environment. We extend the range of validity of the perturbative expansion 
by exponentiating Eq. \eqref{main1} 
\begin{align}
\label{main2}
\left[\rho_S(n)\right]_{ss^\prime} \simeq \exp\bigl[-\,& i\, \alpha (\lambda_s-\lambda_{s^\prime})
g(n) - \alpha^2 (\lambda_s-\lambda_{s^\prime})^2 f(n) \bigr.\nonumber\\
&\bigl. -\, i\, \alpha^2 (\lambda_s^2 - \lambda_{s^\prime}^2) 
\phi(n) \bigr] \left[\rho_S(0)\right]_{ss^\prime},
\end{align}
and our numerics suggest that this expression works well even in strong coupling regime. 
The same approximation was also taken in random matrix study of quantum fidelity 
decay \cite{seligman}, which later justified by supersymmetry calculation 
\cite{stoeckmann}. This approximation is also verified in experimental study  
of fidelity decay in quantum chaotic system \cite{schaefer}. 

The most important term in Eq. \eqref{main2} is the function 
$f(n)$, which is responsible for the decay of the off-diagonal terms of the RDM, leading to 
the loss of coherence of the system. Hence, we call the function $f(n)$ as ``{\it 
decoherence function}". We get rid of the term $g(n)$ by redefining $V_E(l) \equiv V_E(l) - 
\bigl\langle V_E(l) \bigr\rangle$. Then Eq. \eqref{main2} will be exactly identical to the 
expression obtained for the environment consisting of a collection of harmonic oscillators 
\cite{braun1}. However, it does not prove the equivalence of both the expressions. In order 
to do so, we have to show that the decoherence function $f(n)$ is equivalent to the 
decoherence function of Ref. \cite{braun_haake}.

In the following, we closely look at the properties of the decoherence function $f(n)$. From 
the definition of $f(n)$ given in Eq. \eqref{def_fn}, and after doing some algebra, we get:
\begin{equation} 
\label{fn}
f(n) = \frac{1}{2} \sum_{l=1}^{n} \sum_{l^\prime=1}^{n} C_E(l,l^\prime),
\end{equation}  
where $C_E(l,l^\prime)$ is the correlation function of the uncoupled chaotic environment 
given by
\begin{equation}
C_E(l,l^\prime) = \bigl\langle V_E(l) V_E(l^\prime) \bigr\rangle - \bigl\langle V_E(l) 
\bigr\rangle \bigl\langle V_E(l^\prime) \bigr\rangle.
\end{equation}
Here we assume that the environment is strongly chaotic, and the phase space of its 
underlying classical system is bounded. Following Ref. \cite{tanaka}, we assume further 
phenomenological properties of 
$C_E(l,l^\prime)$ : (1) In strongly chaotic regime, the distribution function quickly 
becomes uniform in the phase space, hence $C_E(l,l) = \bigl\langle V_E(l)^2 \bigr\rangle - 
\bigl\langle V_E(l) \bigr\rangle^2$ becomes almost constant within a very short time. So we 
assume $C_E(l,l) \simeq C_0$ for all time. (2) $C_E(l,l^\prime)$ exponentially decays with 
the time-interval $|l-l^\prime|$ with an exponent $\gamma$, i.e., $C_E(l,l^\prime) \simeq 
C_0 \exp(- \gamma |l-l^\prime|)$. Substituting this in Eq. \eqref{fn}, and performing the 
summations we get \cite{tanaka}:
\begin{equation}
\label{pheno_formula}
f(n) \simeq \frac{1}{2} C_0 \left[ n \coth\left(\frac{\gamma}{2}\right) - \frac{1-\exp(-\gamma n)}
{\sinh \gamma -1}\right].
\end{equation} 
Hence, the rate of change of $f(n)$ with time $n$ is :
\begin{equation}   
\frac{\Delta f(n)}{\Delta n} \simeq \frac{1}{2} C_0 \left[ \coth\left(\frac{\gamma}{2}\right) - 
\frac{\gamma \exp(-\gamma n)}{\sinh \gamma -1}\right].
\end{equation}
At the beginning, when $n$ is very small,
\begin{equation}
\frac{\Delta f(n)}{\Delta n} \simeq \frac{1}{2} \left(\frac{C_0 \gamma^2}{\sinh \gamma - 1}
\right)\,n + \mbox{const.}\, , 
\end{equation}
i.e., $f(n)$ evolves quadratically with time ($f(n) \propto n^2$). On the other hand, when 
$n$ is very large,
\begin{equation}
\frac{\Delta f(n)}{\Delta n} \simeq \frac{1}{2} C_0 \coth\left(\frac{\gamma}{2}\right).  
\end{equation}
This suggests long time linear behavior of $f(n)$. Here we identify two distinct 
time-dependent regimes of the decoherence function $f(n)$ : short time quadratic evolution, 
and long time linear evolution. Identical behavior of the decoherence function is reported 
for the the environment consisting of a collection of harmonic oscillators 
\cite{braun_haake}. Thus we finally arrive at our goal, and prove that quantum chaotic 
system and a collection of harmonic oscillators behave equivalently as a model of decohering 
environment.

We now substantiate our results with numerics. Here we consider chaotic kicked top as the 
model of environment. The total Hamiltonian is $H(t) = H_I(t) + H_E(t)$ where $H_I(t)$ is 
the system-environment interaction Hamiltonian, and $H_E(t)$ is the kicked top Hamiltonian. 
As the environment, following version of the kicked top model is used:
\begin{equation}
H_E(t) = \left( \beta J_z + \frac{k}{2j} J_z^2 \right) + \frac{\pi}{2} J_y 
\sum_n \delta(t-n),
\end{equation} 
where $J_i$ is the $i$-th component of the angular momentum operator of the top, $j$ is the 
size of the spin (here, $j=100$), $k$ is the parameter which decides chaoticity in the 
system. The most popular version of the kicked top model does not contain the $\beta$-term. 
We introduce this extra term to remove the parity symmetry $R H(t) R^{-1} = H(t)$, where $R 
= \exp(i \pi J_y)$. The Hilbert space dimension of the kicked top is $N = 2 j + 1 = 201$. We 
set the chaotic parameter at $k=99.0$ which corresponds to classically strongly chaotic 
system, and the parameter $\beta$ at $0.47$. The interaction Hamiltonian between the two 
qubits and the kicked top is:
\begin{equation}
H_I(t) = \alpha h_s \otimes J_z.
\end{equation} 
The parameter $\alpha$ determines the system-environment coupling strength. The system 
consists of two non-interacting qubits, and its coupling agent is $h_s = S_z^{(1)} \otimes 
I^{(2)} + I^{(1)} \otimes S_z^{(2)}$, where $S_z^{(i)} = \sigma_z^{(i)}/2$ and 
$\sigma_z^{(i)}$ is third Pauli matrix. Hence, the computational basis states $\{|00\rangle, 
|01\rangle, |10\rangle, |11\rangle\}$ are also the eigenbasis of $h_s$ with eigenvalues 
$\{-1, 0, 0, 1 \}$. The states $|01\rangle$ and $|10\rangle$ are the degenerate eigenstates 
of $h_s$. Note that, according to Eq. \eqref{main2}, the off-diagonal terms of the system 
state, corresponding to this degenerate subspace, will not decay with time.
 
Decoherence of the system is studied for two different initial states : (1) a Bell state 
$|\phi_s\rangle = \frac{1}{\sqrt{2}}\bigl(|00\rangle - |11\rangle\bigr)$, and (2) a product 
state $|\phi_s\rangle = \frac{1}{\sqrt{2}}\bigl(|0\rangle- |1\rangle\bigr) \otimes 
\frac{1}{\sqrt{2}}\bigl(|0\rangle + |1\rangle\bigr)$. For all numerical studies, a 
generalized $SU(2)$ coherent state \cite{haake_book} is used as initial state for the kicked 
top environment.

\begin{figure}
\includegraphics[width=6.5cm,height=5cm]{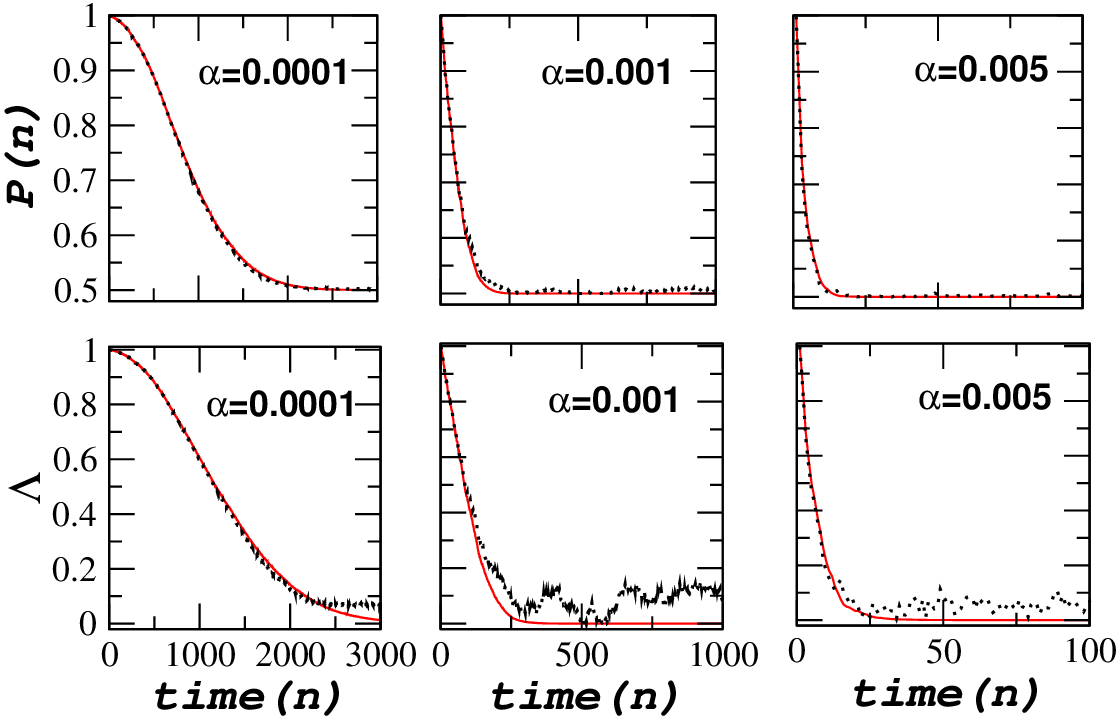}
\caption{(Color online) Evolution of the system purity $P(n)$, and the parameter $\Lambda$ 
determines the entanglement in the system are plotted for three different system-environment 
coupling strengths $\alpha = 0.0001, 0.001, 0.005$. Here the initial state is the Bell state 
(see the text). The dotted line is the exact numerical calculation, whereas the solid line 
is the theoretical estimation. We estimate $P(n)$ and $\Lambda$ from the approximate formula 
of the RDM $\rho_S$ of the system given in Eq. \eqref{main2}.}
\label{pur_conc_bell}
\end{figure}

Fig. \ref{pur_conc_bell} presents the result for the Bell state case. In the computational basis,
diagonal elements of the initial density matrix $\rho_S(0)$ are $(1/2, 0, 0, 1/2)$, and two 
off-diagonal elements have non-zero values : $\langle 00 | \rho_S(0) 
| 11 \rangle = \langle 11 | \rho_S(0) | 00 \rangle = - 1/2$. These off-diagonal elements are
not the part of the degenerate subspace, and therefore they decay in time. So in the 
asymptotic limit : $\rho_S \simeq \mbox{diag}\,(1/2, 0, 0, 1/2)$, the purity $P$ and 
$\Lambda$ of this RDM are $1/2$ and $0$, respectively. In Fig. \ref{pur_conc_bell}, we see 
that the purity $P$ and $\Lambda$ have reached very close to these asymptotic values. The 
rate, at which these two quantities have reached the asymptotic values, is determined by the 
coupling strength $\alpha$. As expected, for weaker coupling the rate is slower than 
stronger coupling. The most important fact is that our approximate formula for the RDM 
derived from the perturbation theory, is not only working well for the weak coupling case 
but also agrees very well for the stronger coupling cases.

\begin{figure}
\includegraphics[width=6.5cm,height=5cm]{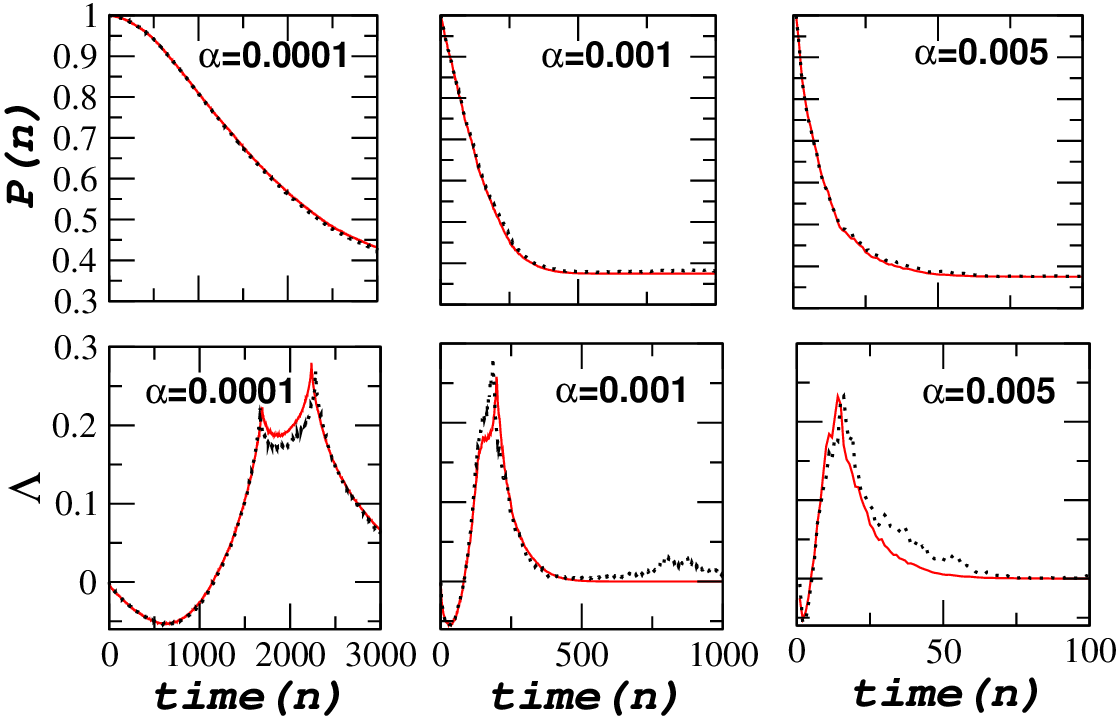}
\caption{(Color online) Evolution of $P(n)$ and $\Lambda$ are plotted. Here initial state is
the product state. Other parameters and descriptions are same as in Fig.
\ref{pur_conc_bell}.}
\label{pur_conc_prod}
\end{figure}

Fig. \ref{pur_conc_prod} presents the result for the product state case. Initially $\Lambda 
= 0$, and then it becomes negative till $n=1142$ for $\alpha = 0.0001$, $n = 81$ for $\alpha 
= 0.001$, and $n = 4$ for $\alpha = 0.005$. Upto this time, entanglement between two qubits 
determined by the concurrence was zero. After that, there is a time interval when $\Lambda > 
0$, and the two qubits become entangled due to their common interaction with the chaotic 
environment. This shows that, like many-body environment \cite{braun}, quantum chaotic 
environment can also generate entanglement between two non-interacting qubits which have no 
entanglement initially. In the asymptotic limit, diagonal
 elements of $\rho_S$ : $(1/4, 1/4, -1/4, -1.4)$, and the off-diagonal elements : $\langle 01 
| \rho_S | 10 \rangle = \langle 10 | \rho_S | 01 \rangle = -1/4$ corresponding to the 
degenerate subspace, survive from decoherence. Therefore, in this limit, the purity $P = 3/8 
= 0.375$, and $\Lambda = 0$. Figure shows again a good agreement between the numerics and 
the theory.

\begin{figure}
\includegraphics[width=5.5cm,height=4cm]{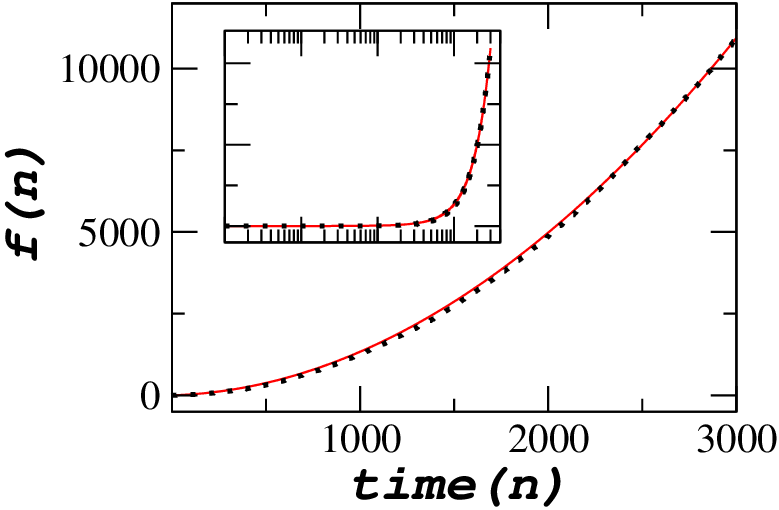}
\caption{(Color online) Evolution of the decoherence function $f(n)$ of the kicked top. The
dotted line is the exact numerics, and the solid line is the theoretical estimation. Inset is
showing the same in semi-logarithmic scale, where the time-axis is in logarithmic scale.}
\label{fn_vs_n}
\end{figure}

Our theoretical formulation establishes the equivalence between quantum chaotic environment 
and many-body environment by showing the equivalence between their decoherence function 
$f(n)$. Therefore, it is very important to investigate $f(n)$ of the kicked top carefully. 
In Fig. \ref{fn_vs_n}, we plot the evolution of the decoherence function of the kicked top. 
Our numerical experiment suggests that the phenomenological formula for $f(n)$ given in Eq. 
\eqref{pheno_formula} is not an exact formula for all time $n$. It describes short time 
quadratic and long time linear behavior well. So instead of fitting the numerics by the 
phenomenological formula, we interpolate it with a simple function $f(n) = a n + b n^2$, a 
combination of a linear and a quadratic term. The interpolation (solid line of Fig. 
\ref{fn_vs_n}) gives $a \simeq 0.1776$ and $b \simeq 1.1561 \times 10^{-3}$. We expect the 
parameter $a$ to be equal to the coefficient of the linear term of Eq. 
\eqref{pheno_formula}, i.e. $\frac{1}{2} C_0 \coth(\gamma/2)$. For the chaotic kicked top, 
$C_0 = \bigl\langle J_z(l)^2 \bigr\rangle - \bigl\langle J_z(l) \bigr\rangle^2 \simeq 1/3$. 
In addition, in strong chaos regime, the parameter $\gamma$ is very large which leads to 
$\coth(\gamma/2) \simeq 1$. Hence, the coefficient of the linear term of Eq. 
\eqref{pheno_formula} is approximately equal to $1/6$, which is very close to the value 
obtained for the parameter $a$, i.e., $a \simeq 0.1776$. Thus we show that the decoherence 
function of the chaotic kicked top has the properties which are similar to the properties of 
any other model of decohering environment.

In conclusion, as a model of decohering environment, we establish the equivalence between 
quantum chaotic system and any other many-body systems. This is an important result, which 
suggests that instead of complex many-body system one can use a simple chaotic system as a 
model of environment. Consequently, analytical investigations of environment induced 
decoherence would be easier, and from numerical point of view as well, this simple model 
requires very less computational resources. We substantiate our analytical formulation with 
strong numerical evidences. These show that quantum chaotic system is good decoherer, and it 
can also create entanglement between two non-interacting particles. The later result is very 
important due to the recent identification of entanglement as a resource for quantum 
computational and informational protocols \cite{quant_comp}.

We thank Drs. A. Tanaka, D. Cohen, and A. Lakshminarayan for useful comments and discussions.

\end{document}